\documentclass[12pt,thmsa]{article}
\usepackage{sw20lart}

\input tcilatex
\begin{document}

\author{Amjad Hussain Shah Gilani \\
National Center for Physics\\
Quaid-i-Azam University\\
Islamabad 45320, Pakistan\\
email: ahgilani@yahoo.com}
\title{How many quarks and leptons ?}
\date{}
\maketitle

\begin{abstract}
There are eight quarks in each family and there are three families of quarks
i.e. $c$, $b$, $t$. Also, we propose similar structure for leptons. The
nature of strong force is named as `third order electroweak'.
\end{abstract}

\section{Introduction}

After more than 30 years of continuous advances we might expect that Quantum
Chromodynamics (QCD), the established theory of strong interactions, would
provide us with a satisfactory understanding of high energy hadronic
reactions \cite{pt2000aug}. Unfortunately this is not yet the case \cite
{hep-ph/0412293}. Data taken by ZEUS\ collaboration at HERA \cite
{hep-ph/9706416} show that the leading particle spectra measured in
photoproduction and in deep inelastic scattering (DIS) (where $Q^2\geq 4$ GeV%
$^2$) are very similar. This suggests that, as pointed out in \cite
{PLB428-383}, QCD hardness scale for particle production in DIS gradually
decreases from a (large) $Q^2$, which is relevant in the photon
fragmentation region, to a soft scale in the proton fragmentation region. We
can therefore expect a similarity of the inclusive spectra of the leading
protons in high energy hadron-proton collisions and in virtual photon-proton
collisions. In other words, we may say that the photon is neither resolving
nor being resolved by the fast emerging protons. This implies that these
reactions are dominated by some non-perturbative mechanism. This is
confirmed by the failure of perturbative QCD \cite{PLB384-388}.

The exploration of physics with $b$-flavoured hadrons offers a very fertile
testing ground for the standard model (SM) description of electroweak
interactions \cite{hep-ph/0003238}. $B$-meson physics is a vast subject,
full of challenges and its decays to light mesons offer the possibility to
access the less well-known entries in the Cabibbo-Kobayashi-Maskawa (CKM)
quark mixing matrix \cite{PRL10-531,PTP49-652} elements, like $V_{ub}$ and $%
V_{ts}$ \cite{hep-ph/9803501}. Therefore, the understanding of flavour
dynamics, and of the related origin of quark and lepton masses and mixings,
is among the most important goals of elementary particle physics. Another
important goal is then to find out whether the Standard Model (SM) is able
to describe the flavour and CP violation observed in nature. All the
existing data on weak decays of hadrons, including rare and CP-violating
decays, can at present be described by the SM within the theoretical and
experimental uncertainities. On the other hand, the SM is an incomplete
theory: some kind of new physics is required in order to understand the
patterns of quark and lepton masses and mixings, and generally to understand
flavour dynamics \cite{hep-ph/0304132}. There are strong theoretical
arguments suggesting that new physics cannot be far from the electroweak
scale.

Radiative $B$ decays to kaons provide a rich laboratory to test the SM and
probe new physics. $B\rightarrow K^{*}\gamma $ is a well established process
among them. Higher resonant kaons such as $K_2^{*}\left( 1430\right) $ are
also measured by CLEO \cite{PRL84-5283} and the $B$ factories \cite
{PRL89-231801,hep-ex/0308021}. Recently, Belle has announced the first
measurement of $K_1\left( 1270\right) $ and upper bound on $K_1\left(
1400\right) $ \cite{hep-ex/0408138}. The higher kaon resonances share lots
of things with $B\rightarrow K^{*}\gamma $. At the quark level, both of them
are governed by $b\rightarrow s\gamma $. All of the accumulated achievements
of $b\rightarrow s\gamma $ can be used in radiative $B$ decays to kaon
resonances. But we have big difference between theory and experiment of
radiative $B$ decays to $K^{*}$ and $K_1$ resonances. The details of
differences are quite opposit \cite{hep-ph/0105302,hep-ph/0409133}. In
short, the form factors 
\begin{eqnarray*}
F_{theory}^{K^{*}} &>&F_{exp}^{K^{*}}, \\
F_{theory}^{K_1} &\ll &F_{exp}^{K_1}.
\end{eqnarray*}
Kown and Lee explained some of the possible candidates of the discrepancy 
\cite{hep-ph/0409133}.

The scheme of the article is: Section 2 is devoted to the existance of QCD.
Quark and lepton families structure is presented in Sec. 3 and 4,
respectively. Conclusion is given in Sec. 5.

\section{Is massless QCD really wrong ?}

Yes, massless QCD is exactly wrong (here ``massless'' means the massless
colored gluons). Why ? QCD is a theory based on the color charges, i.e.,
red, green, and blue. But the value of color charges do not predicted and
these color charges given to the fundamental particles, like quarks and
gluons, on hypothetical basis. It was pointed out for the first time that
the color charges given to gluons violate the group property \cite
{hep-ph/0404026} and only color or anticolor charge can be given to the
gluons. The value of color charges is predicted for the first time in a
recent article by Gilani \cite{hep-ph/0410207}. The quark fractional charges
are also rejected because there was no use of quark fractional charges after
giving the numerical value to the color charges \cite{hep-ph/0410207}. It
was proposed first time that the up-type quarks will carry color charges and
down-type quarks will carry anticolor charges \cite{hep-ph/0410207}. First
time, the gluons structure was proposed by set theory \cite{hep-ph/0404026}.
With the help of set theory, it was suggested that only one gluon is
massless while the remaining seven gluons are massive. Out of seven massive
gluons, three carry color charges and three carry anticolor charges and one
is color singlet. The color singlet gluon is massive than colored or
anticolored gluons. Overall, two gluons are neutral, one is massless and the
other is massive. The remaining six gluons are charged. The mathematical
proof of all the above observations \cite{hep-ph/0404026} about gluons is
given in Ref. \cite{hep-ph/0410207} and obtained exactly the same results as
predicted by set theory \cite{hep-ph/0404026}. The mass of the Higgs boson
is also predicted first time in terms of the mass of $W$-boson \cite
{hep-ph/0410207}.

\section{How many quarks ?}

Question was raised by Gilani \cite{hep-ph/0410207} that if there are three
quarks, then life becomes easy but he did not answer this question
satisfactorily. We pointed out that quarks will not carry fractional charges
but they will carry either color or anticolor charges like the gluons. In
our recent article \cite{hep-ph/0410207}, we proposed the value of color
charges by using the cube roots of unity. By using these value, we give
color charges to up-type quarks i.e. $u$, $c$, $t$ while anticolor charge to
down-type quarks i.e. $d$, $s$, $b$. Here we cannot convince ourselves that
at the same time one type of quarks will carry color charge and the other
type of quarks will carry anticolor charge. What we conclude is, we can only
give color charges to quarks and anticolor charges to antiquarks. If we give
color charge to some of the quarks and anticolor charge to other few quarks
then it will simply be the assumption but there will be a big joke with the
subject and we will go away from the reality. Here we again take the
possibility that there are only three quarks but not six.

We, now, give here a new classifiction to quarks and reject the old
classification by which the quarks were known. We reject the possibility of
six quark families but propose three quark families. We name the three
families as: charm $\left( c\right) $, beauty $\left( b\right) $, and top $%
\left( t\right) $. We give them only color charges. Quarks have also the
same structure as that of gluons which is explained in Refs. \cite
{hep-ph/0404026,hep-ph/0410207}. Three quark families i.e. Charm, Beauty,
top and their quark structure is given in Table 1.

\begin{table}[tbp]
\caption{Three quark families, i.e. Charm $(c)$, Beauty $(b)$, and Top $(t)$%
. }\vspace{0.5cm} 
\begin{tabular}{|l|l|l|l|}
\hline
\textbf{Quarks} & Charm $\left( c\right) $ & Beauty $\left( b\right) $ & Top 
$\left( t\right) $ \\ \hline
Massless & $c^0=d$ & $b^0=u$ & $t^0=s$ \\ \hline
Red colored & $c^r=c^{+1}$ & $b^r=b^{+1}$ & $t^r=t^{+1}$ \\ \hline
Green colored & $c^g=\left( -\frac 12+i\frac{\sqrt{3}}2\right) c^{-1}$ & $%
b^g=\left( -\frac 12+i\frac{\sqrt{3}}2\right) b^{-1}$ & $t^g=\left( -\frac
12+i\frac{\sqrt{3}}2\right) t^{-1}$ \\ \hline
Blue colored & $c^b=\left( -\frac 12-i\frac{\sqrt{3}}2\right) c^{-1}$ & $%
b^b=\left( -\frac 12-i\frac{\sqrt{3}}2\right) b^{-1}$ & $t^b=\left( -\frac
12-i\frac{\sqrt{3}}2\right) t^{-1}$ \\ \hline
$
\begin{array}{c}
\text{Massive} \\ 
\text{colorsinglet}
\end{array}
$ & $c^z$ & $b^z$ & $t^z$ \\ \hline
\textbf{Antiquarks} &  &  &  \\ \hline
Antiblue colored & $\bar{c}^{\bar{b}}=\left( +\frac 12+i\frac{\sqrt{3}}%
2\right) \bar{c}^{+1}$ & $\bar{b}^{\bar{b}}=\left( +\frac 12+i\frac{\sqrt{3}}%
2\right) \bar{b}^{+1}$ & $\bar{t}^{\bar{b}}=\left( +\frac 12+i\frac{\sqrt{3}}%
2\right) \bar{t}^{+1}$ \\ \hline
Antigreen colored & $\bar{c}^{\bar{g}}=\left( +\frac 12-i\frac{\sqrt{3}}%
2\right) \bar{c}^{+1}$ & $\bar{b}^{\bar{g}}=\left( +\frac 12-i\frac{\sqrt{3}}%
2\right) \bar{b}^{+1}$ & $\bar{t}^{\bar{g}}=\left( +\frac 12-i\frac{\sqrt{3}}%
2\right) \bar{t}^{+1}$ \\ \hline
Antired colored & $\bar{c}^{\bar{r}}=\bar{c}^{-1}$ & $\bar{b}^{\bar{r}}=\bar{%
b}^{-1}$ & $\bar{t}^{\bar{r}}=\bar{t}^{-1}$ \\ \hline
\end{tabular}
\end{table}

There must be six quark-Higgs in each family of quarks. We can write the
various mass relations among the quarks on the similar grounds as done for
gluons in Ref. \cite{hep-ph/0410207}.

A stricking question: Is there any relation between the family members of
quarks and/or leptons ?

\section{How many leptons ?}

Three leptons are exist in the literature i.e. electron $\left( e^{-}\right) 
$, muon $\left( \mu ^{-}\right) $, tau $\left( \tau ^{-}\right) $ and their
corresponding neutrinos i.e. electron neutrino $\left( \nu _e\right) $, muon
neutrino $\left( \nu _\mu \right) $, and tau neutrino $\left( \nu _\tau
\right) $. These neutrinos are massless.

Let us assume that leptons have also similar structure as that of gluons 
\cite{hep-ph/0410207} and quarks. If this is so, then we will see that there
will be three families of leptons i.e. electron, muon and tau. The lepton
structure is given in Table 2.

\begin{table}[tbp]
\caption{Three lepton families i.e. electron, muon, tau}\vspace{0.5cm} 
\begin{tabular}{|c|c|c|c|}
\hline
\textbf{Anti-Leptons} & Electron $\left( e\right) $ & Muon $\left( \mu
\right) $ & Tau $\left( \tau \right) $ \\ \hline
$
\begin{array}{c}
\text{Massless} \\ 
\text{neutrino }\left( \nu \right)
\end{array}
$ & $e^0=\nu _e$ & $\mu ^0=\nu _\mu $ & $\tau ^0=\nu _\tau $ \\ \hline
red color & $e^r=e^{+1}$ & $\mu ^r=\mu ^{+1}$ & $\tau ^r=\tau ^{+1}$ \\ 
\hline
green color & $e^g=\left( -\frac 12+i\frac{\sqrt{3}}2\right) e^{-1}$ & $\mu
^g=\left( -\frac 12+i\frac{\sqrt{3}}2\right) \mu ^{-1}$ & $\tau ^g=\left(
-\frac 12+i\frac{\sqrt{3}}2\right) \tau ^{-1}$ \\ \hline
blue color & $e^b=\left( -\frac 12-i\frac{\sqrt{3}}2\right) e^{-1}$ & $\mu
^b=\left( -\frac 12-i\frac{\sqrt{3}}2\right) \mu ^{-1}$ & $\tau ^b=\left(
-\frac 12-i\frac{\sqrt{3}}2\right) \tau ^{-1}$ \\ \hline
$
\begin{array}{c}
\text{Massive colorsinglet} \\ 
\text{(Massive neutrino)}
\end{array}
$ & $e^z=\nu _e^z$ & $\mu ^z=\nu _\mu ^z$ & $\tau ^z=\nu _\tau ^z$ \\ \hline
\textbf{Leptons} &  &  &  \\ \hline
antiblue & $e^{\bar{b}}=\left( +\frac 12+i\frac{\sqrt{3}}2\right) e^{+1}$ & $%
\mu ^{\bar{b}}=\left( +\frac 12+i\frac{\sqrt{3}}2\right) \mu ^{+1}$ & $\tau
^{\bar{b}}=\left( +\frac 12+i\frac{\sqrt{3}}2\right) \tau ^{+1}$ \\ \hline
antigreen & $e^{\bar{g}}=\left( +\frac 12-i\frac{\sqrt{3}}2\right) e^{+1}$ & 
$\mu ^{\bar{g}}=\left( +\frac 12-i\frac{\sqrt{3}}2\right) \mu ^{+1}$ & $\tau
^{\bar{g}}=\left( +\frac 12-i\frac{\sqrt{3}}2\right) \tau ^{+1}$ \\ \hline
antired & $e^{\bar{r}}=e^{-1}$ & $\mu ^{\bar{r}}=\mu ^{-1}$ & $\tau ^{\bar{r}%
}=\tau ^{-1}$ \\ \hline
\end{tabular}
\end{table}

There must be six lepton-Higgs in each family of leptons. We can write the
various mass relations among the quarks on the similar grounds as done for
gluons in Ref. \cite{hep-ph/0410207}.

\section{Conclusions}

We proposed a structure for quarks and leptons similar to the one for gluons 
\cite{hep-ph/0404026,hep-ph/0410207}. We reject the possibility of six quark
flavours and propose that there are only three quark and three lepton
families. The new structure of quark and lepton families are given in Tables
1 and 2.

We summarize here the nature of all the four forces:

\begin{tabular}{|c|c|c|}
\hline
\textbf{Name of force} & \textbf{Nature of force} & \textbf{Gauge bosons} \\ 
\hline
Casimir force & Zeroth order electroweak & $\gamma $ \\ \hline
Gravitational force & First order electroweak & $\gamma ,Z^0$ \\ \hline
Electroweak force & Second order electroweak & $\gamma ,W^{+},W^{-},Z^0$ \\ 
\hline
Strong force & Third order electroweak & Eight gluons \cite
{hep-ph/0404026,hep-ph/0410207} \\ \hline
\end{tabular}

\end{document}